\documentclass[aps,amsfonts,prl,nofootinbib]{revtex4}

\usepackage{epsfig}

\input BoxedEPS
\SetRokickiEPSFSpecial  
\HideDisplacementBoxes

\begin{document}

\title{Quantum Corrections to $Q$-Balls}

\author{N.~Graham\footnote{e-mail:~graham@physics.ucla.edu}}

\affiliation{Department of Physics and Astronomy \\ University of California
at Los Angeles \\ Los Angeles, CA  90095 \\ {\rm hep-th/0105009}}

\begin{abstract}

We extend calculational techniques for static solitons to the case of field
configurations with simple time dependence in order to consider quantum
effects on the stability of $Q$-balls.  These nontopological solitons exist
classically for any fixed value of an unbroken global charge $Q$.  We show
that one-loop quantum effects can destabilize very small $Q$-balls. We show
how the properties of the soliton are reflected in the associated
scattering problem, and find that a good approximation to the full one-loop
quantum energy of a $Q$-ball is given by $\omega - E_0$, where $\omega$ is
the frequency of the classical soliton's time dependence, and $E_0$ is the
energy of the lowest bound state in the associated scattering problem.

{~} 
\end{abstract}

\maketitle

\section{Introduction}

A pure scalar theory in three dimensions with a cubic coupling can support
classically stable, time-dependent non-topological solutions to the
equations of motion carrying an unbroken global charge $Q$, called
$Q$-balls \cite{Coleman}.  Supersymmetric extensions of the standard model
generically contain such objects \cite{SUSY}.  They become particularly
interesting in cosmological applications at small values of $Q$, because
then it is easier for them to form in the early universe \cite{smallq}.  In
this regime, quantum corrections to the soliton's energy become
increasingly important in determining its stability.  The methods of
\cite{method} (see also earlier related work in \cite{methodprev} and
theoretical justification in the Appendices of \cite{dimreg}) provide
an efficient, robust framework for computing quantum corrections to
time-independent field configurations.  In this Letter, we extend this
approach to $Q$-balls.  We show how to express the computation in terms of
an effective time-independent problem.  In this formalism, the full
one-loop quantum correction can be computed efficiently.  We also derive a
very simple estimate for this result.  The result is that we can compare the
energy of the $Q$-ball in the quantum theory to the energy of free particles
carrying the same charge $Q$, and determine if the $Q$-ball remains stable in
the quantum theory.

Our starting point is the classical analysis of $Q$-balls carried out in
\cite{Coleman,smallq}.  We will take the same simple model,
\begin{equation}
{\cal L} =
\frac{1}{2} (\partial_\mu \varphi) (\partial^\mu \varphi) - U(\varphi)
\end{equation}
where $\varphi$ is a complex field with unit charge under a global $U(1)$
symmetry that is unbroken at $\varphi=0$.  We will consider the potential
\begin{equation}
U(\varphi) = \frac{1}{2} M^2 |\varphi|^2 - A |\varphi|^3 +
\lambda |\varphi|^4 \,.
\end{equation}
A particular configuration $\varphi(x,t)$ has charge
\begin{equation}
Q=\frac{1}{2i} \int d^3 x \left(
\varphi^\ast \partial_t \varphi - \varphi \partial_t \varphi^\ast \right) \,.
\end{equation}

Upon fixing the charge $Q$ of the configuration via a Lagrange multiplier
$\omega$, we obtain the classical $Q$-ball solution as the minimum of the
energy functional
\begin{equation}
{\cal E}_\omega[\varphi] =
\int d^3 x \frac{1}{2} |\partial_t \varphi - i\omega \varphi|^2 +
\int d^3 x \left(\frac{1}{2} |\nabla \varphi|^2  + U_\omega(\varphi)\right)
+ \omega Q
\end{equation}
with respect to independent variations of $\varphi(x,t)$ and $\omega$, where
\begin{equation}
U_\omega(\varphi) = U(\varphi) - \frac{1}{2} \omega^2 \varphi^2 \,.
\end{equation}
The $Q$-ball solution then has simple time dependence
\begin{equation}
\varphi(x,t) = e^{i\omega t} \phi(x)
\end{equation}
so we must simply minimize
\begin{equation}
{\cal E}_\omega[\phi] =
\int d^3 x \left( \frac{1}{2} |\nabla \phi|^2  + U_\omega(\phi)\right)
+ \omega Q
\end{equation} by varying $\omega$ and $\phi(x)$. As shown in
\cite{Coleman}, if the quantity $U(\phi)/\phi^2$ is minimized at $\phi_0 >
0$, then for $\omega_0 = \sqrt{2U(\phi_0)/\phi_0^2}$, the effective
potential $U_{\omega_0}(\phi)$ will have degenerate minima.   For
$\omega>\omega_0$, a solution to the equations of motion is given
by the bounce solution for tunneling in three Euclidean dimensions in the
potential $U_\omega(\phi)$.  The bounce is the solution to
\begin{equation}
\frac{d^2}{dr^2} \phi_0(r) + \frac{2}{r} \frac{d}{dr} \phi_0(r) =
U_\omega'(\phi_0(r))
\end{equation}
with the boundary conditions
\begin{equation}
\lim_{r\to\infty} \phi_0(r) = 0 \hbox{~~~~and~~~~}
\left. \frac{d}{dr} \phi_0(r) \right|_{r=0} = 0 \,.
\end{equation}
We can then find the solution using the shooting method detailed in
\cite{Colesym}, and minimize the resulting energy over $\omega$.  For large
enough $Q$, the optimal value of $\omega$ approaches $\omega_0$, allowing
\cite{Coleman} to use the thin-wall approximation to demonstrate the
existence of a global minimum, which is the $Q$-ball solution.  For small
$Q$, the optimal value of $\omega$ approaches $M$, and \cite{smallq} uses
the thick-wall approximation to show that there exists a global minimum in
this case as well.  Thus classically bound solitons exist all the way down
to $Q=1$.  For all $Q$, we have $\omega_0 < \omega < M$ at the minimum.
We can thus consider the classical binding energy as a function of $Q$ by
comparing the $Q$-ball's energy to $QM$, the energy of a collection of free
particles carrying charge $Q$.

Although $Q$-balls are classically stable even as $Q\to 1$, the binding energy
per charge is going to zero in this limit.  This case is of particular interest
for cosmological applications, however, since $Q$-balls of large charge,
while favored energetically, are disfavored as the temperature increases by
their low entropy.  To answer the question of whether $Q$-balls have a
significant chance of being formed in the early universe, we must therefore
verify that the classical conclusions are not invalidated by quantum
corrections.

\section{Quantum Corrections}

To compute the leading quantum correction the $Q$-ball energy, we extend
the method of \cite{method}.  We write the quantum field $\varphi$ as the
classical solution plus a quantum correction, which we write in corotating
coordinates
\begin{equation}
\varphi(x,t) = e^{i\omega t} \left(\phi_0(x) + \eta(x,t)\right)
\end{equation}
where we can then expand the quantum field $\eta(x,t)$ in small
oscillations, which are given by the solutions to 
\begin{equation}
\left[(\partial_t - i\omega)^2 - \nabla^2 + U''(\phi_0(x))\right]\psi(x,t) = 0
\,.
\end{equation}
Parametrizing $\psi(x,t) = e^{i\omega t} e^{-iEt} \psi(x)$ gives the mode an
energy $E - \omega$, where the time-independent wavefunction $\psi(x)$ solves
\begin{equation}
\left[-\nabla^2 + U''(\phi_0(x))\right]\psi(x) = E^2 \psi(x)
\label{smosc}
\end{equation}
which is an ordinary Schr\"odinger equation. The Casimir energy is given
formally by the sum over zero-point energies of these oscillations
\begin{equation}
{\cal E_C^{\mbox{\tiny bare}}}[\phi_0] \sim \frac{1}{2}
\sum_j |E_j - \omega| \,.
\label{modesum1}
\end{equation}
Since the spectrum of eq.~(\ref{smosc}) is symmetric in $E\to -E$, we can
sum over both signs of the energy and obtain
\begin{equation}
{\cal E_C^{\mbox{\tiny bare}}}[\phi_0] \sim \frac{1}{2} \sum_{E_j\geq 0}
\left( |E_j + \omega| + |E_j - \omega| \right) =
\sum_{E_j\geq 0} \mbox{max} \left(|\omega |, |E_j| \right) \,.
\label{modesum2}
\end{equation}

We will use the methods of \cite{method} to extract the quantum correction
to the energy in terms of the continuum scattering data for the reduced
problem of eq.~(\ref{smosc}).  Since the potential is spherically symmetric,
we can decompose the spectrum into partial waves $\ell$.  We have
wavefunctions
\begin{equation}
\psi_\ell(x) = \frac{Y_{\ell m}(\Omega)}{r} \eta_\ell(r)
\end{equation}
for $m=-\ell,-\ell+1, \dots,\ell-1, \ell$.  The radial wavefunction
$\eta_\ell(r)$ satisfies
\begin{equation}
\left(-\frac{d^2}{dr^2} + \frac{\ell(\ell+1)}{r^2}
+U''(\phi_0(r))\right) \eta_\ell(r) = E^2 \eta_\ell(r)
\label{smoscred}
\end{equation}
with scattering boundary conditions.

In each partial wave, we will find a continuum starting at $E=M$ and
possibly bound states with $0 \leq E_j \leq M$.  (Since the spectrum is
symmetric in $E$, we will only consider $E\geq 0$.)  It is instructive to
consider the properties of eq.~(\ref{smosc}) that betray its origin from a
field theory soliton.  The full oscillation spectrum should have a zero
mode in the $\ell = 1$ channel, corresponding to the translation invariance
of the $Q$-ball solution.  The threefold degeneracy of this state
corresponds to the three directions of translation.  From
eq.~(\ref{modesum1}), we see that in the reduced problem, the zero mode
appears as a bound state with energy $E=\omega$.  Since this state appears
in the $\ell=1$ channel, there must exist an even more tightly bound state
in the $\ell=0$ channel.  In the case of an ordinary static solution, this
state would correspond to an instability of the full soliton.  But from
eq.~(\ref{modesum2}), we see that the destabilizing effect of this mode is
neutralized by the time dependence of the classical solution, which results
in it making the same contribution to the mode sum as the zero modes do. 
All other modes have energies greater than $\omega$.

Having rewritten the Casimir energy in terms of the eigenmodes of the
reduced scattering problem in eqs.~(\ref{smosc}) and (\ref{smoscred}), we
are prepared to apply the methods of \cite{method}.  We obtain the
renormalized Casimir energy as a sum over partial waves $\ell$.  In each
partial wave we have a sum over positive energy bound states $E_{j,\ell}$
and an integral over continuum states weighted by the density of states
$\rho(k)$, where $k = \sqrt{E^2 - M^2}$. We subtract the corresponding
integral in the free case, obtaining
\begin{equation}
{\cal E_C}^{\mbox{\tiny bare}}[\phi_0] =
\sum_{\ell=0}^\infty (2\ell + 1) \left[ \int_0^\infty \frac{dk}{\pi} E
\left(\rho_\ell(k) - \rho^0_\ell(k) \right)
+ \sum_j \mbox{max}(\omega, E_{j,\ell}) \right] \,.
\label{fullcas0}
\end{equation}
This integral diverges in the unrenormalized theory, because we have not
yet included the contribution of the counterterms.  We will compute the
continuum integral by relating the difference between the interacting and
free densitites of states to the phase shifts,
\begin{equation}
\rho_\ell(k) - \rho^0_\ell(k) = \frac{1}{\pi} \frac{d}{dk} \delta_\ell(k) \,.
\end{equation}
The Born expansion for the phase shift is then in exact correspondence with
the expansion of the effective energy in one-loop diagrams with all
possible insertions of the background field.  Subtracting the first two
Born approximations from the phase shift corresponds to subtracting the
first two diagrams in this expansion, which are the only divergent terms. 
The remaining integral is then finite.  We then add the divergent terms
back in, together with counterterms, as ordinary Feynman diagrams.
Full details are given in \cite{method}.

We thus obtain the renormalized Casimir energy
\begin{equation}
{\cal E_C}[\phi_0] = \Gamma^{(2)}[\phi_0]
+ \sum_{\ell=0}^\infty (2\ell + 1)
\left[ \int_0^\infty \frac{dk}{\pi} E \frac{d}{dk}
\left(\delta_\ell(k) - \delta^{(1)}_\ell(k) - \delta^{(2)}_\ell(k) \right)
+ \sum_j \mbox{max}(\omega, E_{j,\ell}) \right] \,.
\label{fullcas}
\end{equation}
where $\delta_\ell(k)$ is the scattering phase shift in partial wave
$\ell$, $\delta^{(1)}_\ell(k)$ and $\delta^{(2)}_\ell(k)$ are its first and
second Born approximations, and $\Gamma^{(2)}[\phi_0]$ is the contribution
to the energy from the two-point function, computed in ordinary Feynman
perturbation theory.  This piece includes the counterterms, which we fix
using physical renormalization conditions.  We demand that the tadpole
graph vanish, and that the mass of the free $\varphi$ particle is
unchanged. In addition to holding the location of the pole  in the
propagator fixed at $M$, we also perform wavefunction renormalization so
that its residue is unchanged as well.

We can now compute eq.~(\ref{fullcas}) directly.  We simply require the
scattering phase shifts, their Born approximations, and the bound states of
eq.~(\ref{smosc}).  Efficient algorithms for obtaining these are detailed
in \cite{method}.  The contribution to the energy from the two-point
function is computed using conventional techniques, giving
\begin{equation}
\Gamma_2 [\phi_0] = \int_0^\infty\frac {4q^2dq}{(4\pi)^4}
\left[ \left(2 \frac{\sqrt{q^2+4 M^2}}{q} {\rm arctanh}
\frac{q}{\sqrt{q^2+4M^2}} - \frac{5\pi}{3\sqrt{3}} + 1 \right)
|\tilde \chi (q)|^2 - 4 q^2 \left(\frac{2\pi}{3\sqrt{3}} - 1\right)
|\tilde{\phi}(q)|^2 \right]
\end{equation}
where $\tilde\chi(q)$ and $\tilde\phi(q)$ are the spatial Fourier
transforms of $U''(\phi_0(r))-M^2$ and $\phi(r)$ respectively.

Examining this calculation in detail yields a very accurate estimate for
the quantum correction to the energy, which is very easy to compute.  Using
the analysis of bound states above, we can separate eq.~(\ref{fullcas}) into
\begin{equation}
{\cal E_C}[\phi_0] = \Gamma^{(2)}[\phi_0]
+ \sum_{\ell=0}^\infty (2\ell + 1)
\left[ \int_0^\infty \frac{dk}{\pi} E \frac{d}{dk}
\left(\delta_\ell(k) - \delta^{(1)}_\ell(k) - \delta^{(2)}_\ell(k) \right)
+ \sum_j E_{j,\ell} \right] + (\omega - E_0)
\label{fullcassep}
\end{equation}
where $E_0$ is energy of the most tightly bound state, which appears in the
$\ell=0$ channel.  It is the only state with energy less than $\omega$.  We
define the reduced Casimir energy as
\begin{eqnarray}
{\cal E_C^{\mbox{\tiny reduced}}}[\phi_0]
&=& \Gamma^{(2)}[\phi_0] + \sum_{\ell=0}^\infty (2\ell + 1)
\left[ \int_0^\infty \frac{dk}{\pi} E \frac{d}{dk}
\left(\delta_\ell(k) - \delta^{(1)}_\ell(k) - \delta^{(2)}_\ell(k) \right)
+ \sum_j E_{j,\ell} \right] \cr
&=& \Gamma^{(2)}[\phi_0] + \sum_{\ell=0}^\infty (2\ell + 1)
\left[ \int_0^\infty \frac{dk}{\pi} (E-M) \frac{d}{dk}
\left(\delta_\ell(k) - \delta^{(1)}_\ell(k) - \delta^{(2)}_\ell(k) \right)
+ \sum_j (E_{j,\ell}-M) \right]
\label{fullcasred}
\end{eqnarray} where we have used Levinson's theorem in the second line. 
This quantity is simply the Casimir energy of a time-independent soliton
giving rise to the reduced small oscillations of eq.~(\ref{smosc}).  (Of
course, such a soliton would not solve the field theory equations of
motion, but we could imagine holding it in place with an external source).
The reduced potential is shallow and slowly varying, especially in the
limit of small $Q$, which corresponds to $\omega$ approaching $M$.  It
causes only a slight deformation of the small oscillations spectrum --- in
particular, there is only one state bound more tightly than $\omega$.  For
a generic potential of this kind, the contributions from the bound states
and continuum will be opposite in sign; roughly, rearrangement of the
continuum spectrum partially compensates for the effect of the states that
become bound.

A direct application of \cite{method} allows us to evaluate the full
result of eq.~(\ref{fullcasred}).  We can also estimate this result in the
derivative expansion.  To lowest order, we have simply the effective
potential contribution
\begin{equation}
{\cal E_C^{\mbox{\tiny reduced,DE}}}[\phi_0]
= \int_0^\infty \frac{r^2 dr}{8\pi}
M^4\left((1+z)^2 \log (1+z) - z - \frac{3}{2} z^2\right)
\end{equation}
where
\begin{equation}
z = \frac{U''(\phi_0(r)) - M^2}{M^2} \,.
\end{equation}

Using either technique, explicit computations show that this reduced
Casimir energy is very small compared to the classical binding energy of
the $Q$-ball (typically 5\% or less for small $Q$). Thus we lose very little
accuracy by dropping this term, obtaining a very simple estimate for the
Casimir energy:
\begin{equation}
{\cal E_C}[\phi_0] \approx {\cal E_C^{\mbox{\tiny est}}}[\phi_0] =
\omega - E_0 \,.
\label{casest}
\end{equation}

\section{Applications}

\begin{figure}[hbt]
\centerline{
\BoxedEPSF{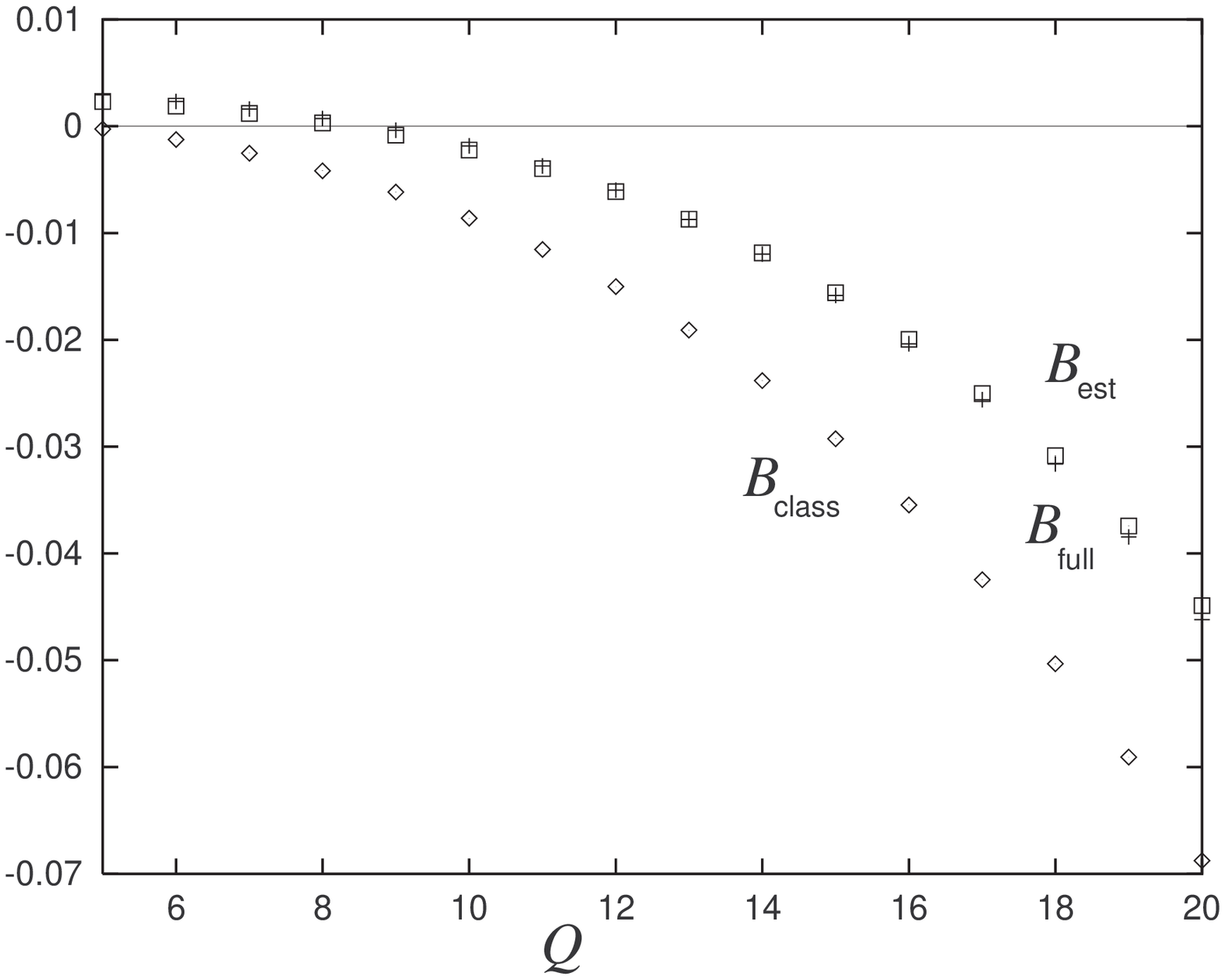 scaled 400} \hfil
\BoxedEPSF{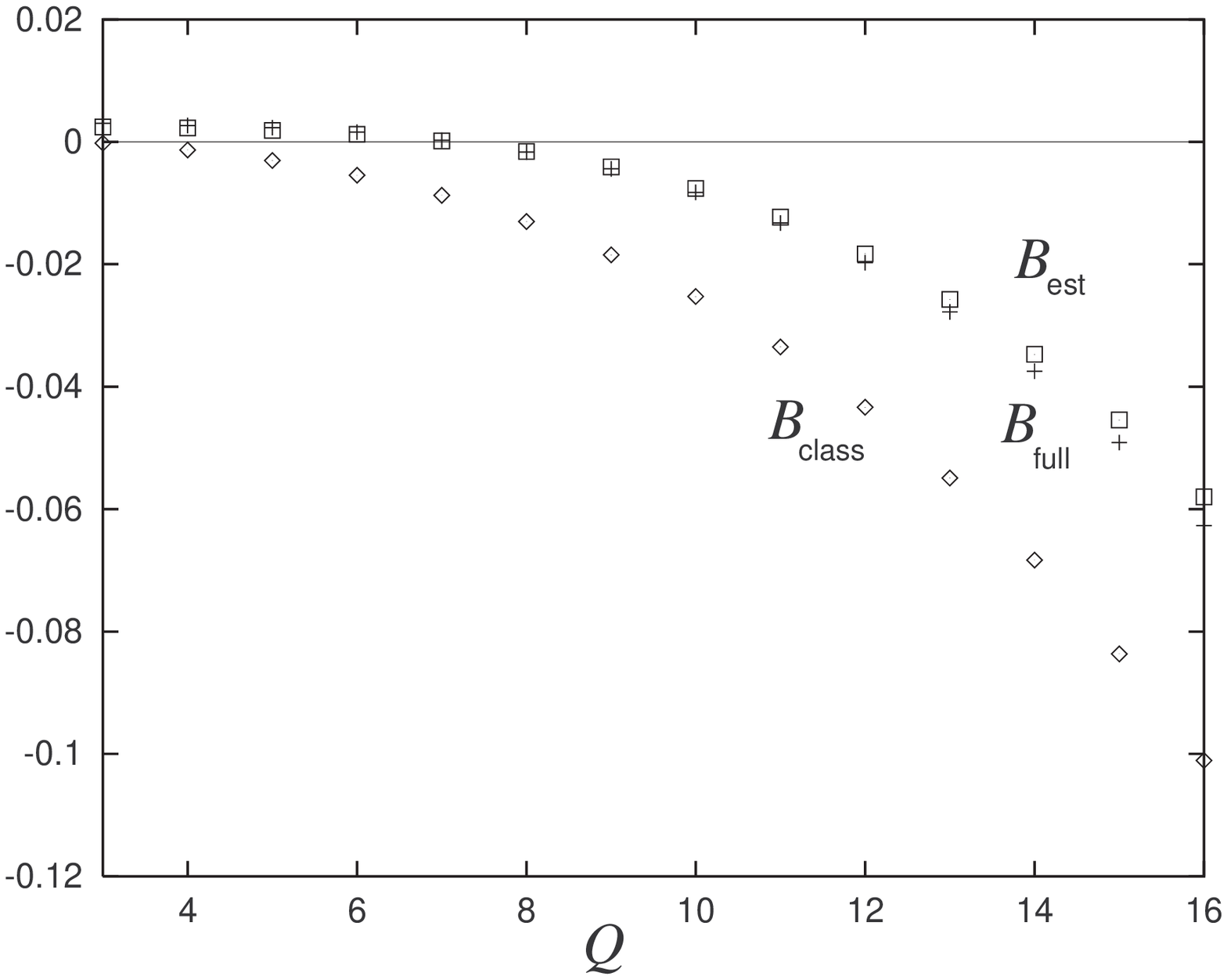 scaled 400}}
\caption{\sl
$Q$-ball binding as a function of $Q$, in units of $M$.  Parameters are
$A=0.325M$ and $\lambda = 0.055$ (left panel), and
$A=0.425M$ and $\lambda = 0.095$ (right panel).  Shown are three
calculations of the difference ${\cal B}$ between the energy of a $Q$-ball
and the energy of a state with charge $Q$ built on the trivial vacuum: the
classical approximation ${\cal B_{\mbox{\tiny class}}}[\phi_0]$, the full
one-loop calculation ${\cal B_{\mbox{\tiny full}}}[\phi_0]$, and the
estimated one-loop result ${\cal B_{\mbox{\tiny est}}}[\phi_0]$.}
\label{Qfig}
\end{figure}

To see whether the $Q$-ball is stable, we must compare its energy to the
energy of a state with the same charge built on the trivial vacuum
\begin{equation}
{\cal B}[\phi_0] = {\cal E}[\phi_0] - QM \,.
\end{equation}

Figure~\ref{Qfig} shows the result of different calculations of ${\cal E}$,
each as a function of $Q$, for two choices of the coupling constants.  In
both cases, they are chosen so that $\varphi = 0$ remains the global minimum
of $U(\varphi)$.\footnote{I thank M.~Postma for reminding me of this
requirement.}  We work in units of $M$, which sets the scale of the problem.
In the classical approximation,
\begin{equation}
{\cal B_{\mbox{\tiny class}}}[\phi_0] = {\cal E}_\omega[\phi_0] - QM
\end{equation}
we see the result of \cite{smallq}:  the $Q$-ball is stable for all $Q$,
though the binding energy per charge is going to zero as $Q\to 0$.  In the
full one-loop calculation,
\begin{equation}
{\cal B_{\mbox{\tiny full}}}[\phi_0] = {\cal E}_\omega[\phi_0] + {\cal
E_C}[\phi_0] - QM
\end{equation}
we see that the quantum corrections overwhelm the weak classical binding up
to $Q_{\mbox{\tiny min}}\approx 7$.  Above this value, the $Q$-ball is stable. 
Finally, we see that using eq.~(\ref{casest}) to approximate to the one-loop
result by taking
\begin{equation}
{\cal B_{\mbox{\tiny est}}}[\phi_0] = {\cal E}_\omega[\phi_0] +
{\cal E_C^{\mbox{\tiny est}}}[\phi_0] - QM
\end{equation}
yields a result that is very close the full one-loop result.  It is also
interesting to note that this approximation is particularly good near the
value of $Q$ at which the $Q$-ball becomes bound in the full one-loop
calculation.

\section{Conclusions}

We have computed one-loop quantum corrections to the energies of
$Q$-balls.  For small $Q$, these corrections can play an important role in
determining the stability of these extended objects.  Since $Q$-balls are
genuine solutions to the classical equations of motion, and we are working
in a regime where the coupling constants are small, the one-loop
approximation should be very good.  We have seen that the one-loop
correction is indeed very small compared to the classical energy of the
$Q$-ball, although can be significant when compared to its binding energy,
the difference between the energy of the $Q$-ball and the energy of a
collection of free particles carrying the same charge.  The higher-loop
corrections should be correspondingly smaller than the one-loop
corrections, and therefore can safely be neglected.  Of course, if the
theory contains other particles coupled to the $\varphi$ field, we may need
to include their contributions as well, using the same techniques as we
have developed here.  For small coupling constants, the one-loop
approximation should continue to be reliable.  Within the one-loop
approximation, we have seen that the quantum correction can be very
accurately estimated at small $Q$ by considering the difference between the
Lagrange multiplier $\omega$, which gives the frequency of the $Q$-ball's
time dependence, and the energy $E_0$ of the lowest bound state of the
small oscillations potential $U''(\phi_0(x))$.  The result is a prediction
of the minimum value of $Q$ for which the $Q$-ball is stable in the quantum
theory.  For typical values of the coupling constants, we find
$Q_{\mbox{\tiny min}} \approx 7$.

\section{Acknowledgments}

I would like to thank A.~Kusenko for helpful discussions, suggestions, and
references at all stages in this project.  This work is supported by the
U.S.~Department of Energy (D.O.E.) under cooperative research agreement
\#DE-FG03-91ER40662.

\end{document}